\documentclass[aps,pra,floatfix,reprint,showpacs,superscriptaddress,groupedaddress, nofootinbib]{revtex4-1}

\usepackage{latexsym}
\usepackage{graphicx}
\usepackage{newtxmath}
\usepackage[colorlinks]{hyperref}
\usepackage{rotating}
\usepackage[normalem]{ulem}
\usepackage{hyperref}
\usepackage{amsmath,amsfonts}
\usepackage{mathtools}
\usepackage{bm}
\usepackage{color}
\usepackage{textgreek}
\usepackage{soul}  
\usepackage{qcircuit}
\usepackage[greek,english]{babel}

\newcommand{\ket}[1]{\ensuremath{| #1 \rangle}}

\usepackage{color}

\newcommand{\lyxmathsym}[1]{\ifmmode\begingroup\def\b@ld{bold}
  \text{\ifx\math@version\b@ld\bfseries\fi#1}\endgroup\else#1\fi}

\usepackage{amsfonts}
\usepackage{mathrsfs}
\usepackage{physics}
\usepackage{multirow}

\usepackage{tikz}
\usepackage{qcircuit}
\usepackage{tikzit}
\usepackage{braket}
\usepackage{caption}
\usepackage{subcaption}
\input{config/zx.tikzdefs}

\tikzstyle{box}=[shape=rectangle, text height=1.5ex, text depth=0.25ex, yshift=0.5mm, fill=white, draw=black, minimum height=5mm, yshift=-0.5mm, minimum width=5mm, font={\small}]
\tikzstyle{Z dot}=[inner sep=0mm, minimum size=2mm, shape=circle, draw=black, fill={rgb,255: red,216; green,248; blue,216}, tikzit fill={rgb,255: red,216; green,248; blue,216}]
\tikzstyle{Z phase dot}=[minimum size=4.75mm, font={\footnotesize}, shape=rectangle, rounded corners=1.9mm, inner sep=0.1mm, outer sep=-2mm, scale=0.8, tikzit shape=circle, draw=black, fill={rgb,255: red,216; green,248; blue,216}, tikzit draw=blue, tikzit fill={rgb,255: red,216; green,248; blue,216}]
\tikzstyle{X dot}=[Z dot, shape=circle, draw=black, fill={rgb,255: red,232; green,165; blue,165}, tikzit fill={rgb,255: red,232; green,165; blue,165}]
\tikzstyle{X phase dot}=[Z phase dot, tikzit shape=circle, tikzit fill={rgb,255: red,232; green,165; blue,165}, fill={rgb,255: red,232; green,165; blue,165}, font={\footnotesize}, tikzit draw=blue]
\tikzstyle{XD dot}=[shape=XDdot, inner sep=2pt, draw=black]
\tikzstyle{XD phase dot}=[shape=XDdotphase, minimum size=4.75mm, font={\footnotesize}, inner sep=.1mm, outer sep=0mm, scale=0.8, tikzit shape=circle, rounded corners=1.9mm, draw=black]
\tikzstyle{zn}=[shape=zn, tikzit draw=black, draw=black, inner sep=2pt]
\tikzstyle{hadamard}=[fill={rgb,255: red,140; green,220; blue,248}, draw=black, shape=rectangle, inner sep=0.6mm, minimum height=1.5mm, minimum width=1.5mm, xslant=0.5]
\tikzstyle{hz}=[hadamard, fill={rgb,255: red,216; green,248; blue,216}, shape=rectangle, tikzit fill={rgb,255: red,216; green,248; blue,216}, minimum height=2 mm, minimum width=1.25 mm, tikzit draw=black]
\tikzstyle{hx}=[hadamard, fill={rgb,255: red,232; green,165; blue,165}, shape=rectangle, tikzit fill={rgb,255: red,232; green,165; blue,165}, minimum height=2 mm, minimum width=1.25 mm, tikzit draw=black]
\tikzstyle{vertex}=[inner sep=0mm, minimum size=1mm, shape=circle, draw=black, fill=black]
\tikzstyle{vertex set}=[inner sep=0mm, minimum size=1mm, shape=circle, draw=black, fill=white, font={\footnotesize\boldmath}]
\tikzstyle{meter}=[draw, fill=white, minimum width=2em, minimum height=1.5em, rectangle, path picture={\draw ([shift={(.1,.24)}]path picture bounding box.south west) to[bend left=50] ([shift={(-.1,.24)}]path picture bounding box.south east);\draw[-{Latex[scale=0.6]}] ([shift={(0,.1)}]path picture bounding box.south) -- ([shift={(.3,-.1)}]path picture bounding box.north);}, tikzit shape=rectangle]
\tikzstyle{white dot}=[Z dot]
\tikzstyle{gray dot}=[X dot]
\tikzstyle{white phase dot}=[Z phase dot]
\tikzstyle{gray phase dot}=[X phase dot]
\tikzstyle{red ket}=[fill={rgb,255: red,232; green,165; blue,165}, draw=black, shape=isosceles triangle, tikzit fill={rgb,255: red,232; green,165; blue,165}, tikzit draw=black, inner sep=0 mm, outer sep=2 mm]
\tikzstyle{tiny none}=[none, font={\tiny}]
\tikzstyle{green ket}=[fill={rgb,255: red,216; green,248; blue,216}, draw=black, shape=isosceles triangle, tikzit fill={rgb,255: red,216; green,248; blue,216}, tikzit draw=black, inner sep=0 mm, outer sep=2 mm]
\tikzstyle{filament}=[hadamard, fill=yellow, draw=none, minimum height=0.01mm]
\tikzstyle{unit circle}=[shape=circle, minimum size=42.5 mm, fill=none, draw={rgb,255: red,223; green,223; blue,223}, tikzit draw={rgb,255: red,223; green,223; blue,223}]
\tikzstyle{new style 0}=[shape=ellipse, minimum height=10 mm, minimum width=100 mm, fill=white, draw=black]
\tikzstyle{gate}=[box, minimum height=10mm, minimum width=10mm]
\tikzstyle{2 control}=[vertex set, draw=blue, inner sep=0.5pt]
\tikzstyle{1 control}=[2 control, draw=red]
\tikzstyle{0 control}=[2 control, draw=black]
\tikzstyle{small dot}=[vertex, minimum size=1 mm, draw=black, tikzit draw=black, tikzit fill=black, tikzit shape=circle]
\tikzstyle{tallbox}=[box, minimum height=12mm]
\tikzstyle{targ}=[vertex set, minimum size=0.5mm, inner sep=-0.5mm, tikzit shape=circle, shape=circle, tikzit draw=black]
\tikzstyle{hadamardbox}=[hadamard, xslant=0]

\tikzstyle{directedarrow}=[draw={rgb,255: red,223; green,223; blue,223}, ->, tikzit draw={rgb,255: red,223; green,223; blue,223}, line width=1 pt]
\tikzstyle{simple}=[-]
\tikzstyle{hadamard edge}=[-, color={rgb,255: red,0; green,100; blue,248}, dashed, dash pattern=on 2pt off 0.7pt, tikzit draw={rgb,255: red,0; green,100; blue,248}]
\tikzstyle{brace edge}=[-, tikzit draw=blue, decorate, decoration={brace,amplitude=1mm,raise=-1mm}]
\tikzstyle{gray}=[-, draw={rgb,255: red,223; green,223; blue,223}, line width=1 pt]
\tikzstyle{arrow}=[<-, draw={rgb,255: red,128; green,128; blue,128}]
\tikzstyle{double-arrow}=[draw={rgb,255: red,128; green,128; blue,128}, <->]
\tikzstyle{dashed edge}=[-, dashed, dash pattern=on 2pt off 0.5pt, draw=black]
\tikzstyle{diredge}=[->]
\tikzstyle{double edge}=[-, double, shorten <=-1mm, shorten >=-1mm, double distance=2pt]
\tikzstyle{thin}=[-, line width=0.05mm]
\tikzstyle{thin gray}=[-, draw={rgb,255: red,223; green,223; blue,223}, line width=0.05mm]
\tikzstyle{less thin}=[-, line width=0.1mm]
\tikzstyle{dashed gray edge}=[-, dashed edge, draw={rgb,255: red,128; green,128; blue,128}]
\tikzstyle{light right directed arrow}=[->, directedarrow, draw={rgb,255: red,223; green,223; blue,223}, line width=0.2mm]
\tikzstyle{diredge0.3}=[->, line width=0.3 mm]
\tikzstyle{less thin gray}=[-, draw={rgb,255: red,223; green,223; blue,223}]
\tikzstyle{dashed thin purple}=[-, dashed, line width=0.1mm, draw={rgb,255: red,128; green,106; blue,219}]
\tikzstyle{hadamardedge}=[-, color={rgb,255: red,100; green,200; blue,248}, dashed, dash pattern=on 2pt off 0.7pt, tikzit draw={rgb,255: red,120; green,220; blue,248}]

\hypersetup{colorlinks=true,urlcolor=blue,linkcolor=blue}

\begin{document}
\title{Encoding optimization for quantum machine learning demonstrated on a superconducting transmon qutrit} 

\author{Shuxiang Cao$^{1}$}
\thanks{These two authors contributed equally}
\author{Weixi Zhang$^{2}$}
\thanks{These two authors contributed equally}
\author{Jules Tilly$^{3}$}
\author{Abhishek Agarwal$^{2}$}
\author{Mustafa Bakr$^{1}$}
\author{Giulio Campanaro$^{1}$}
 \author{Simone D Fasciati$^{1}$}
 \author{James Wills$^{1}$}
 \author{Boris Shteynas$^{1}$}
 \author{Vivek Chidambaram$^{1}$}
\author{Peter Leek$^{1}$}
\author{Ivan Rungger$^{2}$}
\email{ivan.rungger@npl.co.uk}
\affiliation{$^1$ Clarendon Laboratory, Department of Physics, University of Oxford, Oxford, OX1 3PU, United Kingdom}
\affiliation{$^2$ National  Physical  Laboratory,  Teddington,  TW11  0LW,  United  Kingdom}
\affiliation{$^3$ InstaDeep, London, W2 1AY, United Kingdom}

\begin{abstract}
Qutrits, three-level quantum systems, have the advantage of potentially requiring fewer components than the typically used two-level qubits to construct equivalent quantum circuits. This work investigates the potential of qutrit parametric circuits in machine learning classification applications. We propose and evaluate different data-encoding schemes for qutrits, and find that the classification accuracy varies significantly depending on the used encoding. We therefore propose a training method for encoding optimization that allows to consistently achieve high classification accuracy. Our theoretical analysis and numerical simulations indicate that the qutrit classifier can achieve high classification accuracy using fewer components than a comparable qubit system. We showcase the qutrit classification using the optimized encoding method on a superconducting transmon qutrit, demonstrating the practicality of the proposed method on noisy hardware. Our work demonstrates high-precision ternary classification using fewer circuit elements, establishing qutrit parametric quantum circuits as a viable and efficient tool for quantum machine learning applications.
\end{abstract}
\maketitle

\section{Introduction}

Lying at the intersection between quantum computation and machine learning (ML), quantum machine learning (QML) has become a field of growing interest in recent years \cite{QML_review, Schuld2014, adcock2015advances, Schuld2018, garg2020advances, Zhang2020, Ciliberto2018, li2021recent, ganguly2021quantum}. QML encompasses a large number of techniques and learning tasks, including variational quantum optimization problems \cite{cerezo2020variational}, resolution of molecular electronic structure problems \cite{peruzzo2014variational, McClean2016, tilly2021variational}, and more general ML tasks, such as using quantum neural networks (QNN) to perform a classification task on classical data \cite{Sahni2006, Quantum_classifier, Iris_multiclass, DQNN, QNN, Schuld2020, Potempa2021, QuestQNN}. 

QNNs acting on classical input data require the definition of a process to encode the classical information into quantum information \cite{Schuld2019, Havlek2019}. A parameterized quantum circuit is then used as ansatz for the classifier, and its parameters are optimized to minimize a loss function, in analogy to classical neural networks \cite{Benedetti2019}. The most common quantum circuits used as QNNs are tree tensor networks (TTN) \cite{TTN}, multi-scale entanglement renormalization ans\"atze (MERA) \cite{MERA, Cincio2008, Evenbly2010}, and their extensions, which include a quantum version of convolutional neural networks (QCNN) \cite{Cong2019}. 
While the cost of encoding classical data onto quantum states in terms of the required circuit depth can limit QML in its potential to bring an advantage over classical methods \cite{Aaronson2015, Ciliberto2018}, there are indications that QML may be able to perform tasks that require degrees of entanglement which would be intractable for classical ML \cite{Benedetti2019, Quantum_classifier}.

Efficient encoding of classical data into quantum data is understood to be the likely critical driver for the advantage of QML over classical ML \cite{Schuld2019, Havlek2019, Huang2021}. One can decide to encode the data onto the amplitudes of the wave function of the quantum state, encoding a vector of $\mathcal{O}(2^N)$ data point onto $N$ qubits. This may however require exponential circuit depth to implement, negating any advantage brought from using quantum computing \cite{Aaronson2015}. To reduce the encoding circuit depth at the expense of using more qubits, one can use an encoding scheme based on an unentangled product state of $N$ qubits, with the amplitude of the wavefunction coefficient of each qubit encoding a single data point, allowing to encode a vector of $N$ data points with a circuit depth of $\mathcal{O}(1)$. We refer to this method as non-entangled amplitude encoding (NAE). Alternative methods have also been suggested to train a parameterized circuit to learn the most appropriate encoding \cite{lloyd2020quantum} instead of using a fixed circuit. 

Current quantum computers generally operate with qubits, which are two-level quantum systems, the quantum equivalent of the binary bit used in classical computing. One can also consider the use of a general \textit{d}-level quantum system, with \textit{d} being an arbitrary integer and the corresponding unit of quantum information being denoted a qudit. For a review of quantum computation using qudits, we direct readers to Ref. \cite{Qudit_review}. The 3-level quantum system is referred to as qutrit. Having access to more than two states allows one to perform specific tasks with a quantum circuit that requires fewer qudits and gates compared to an equivalent qubit circuit \cite{Qudit_gates, Luo2014, Universal_qudit_computation}. For example, qutrits enable simplification of the Toffoli gate \cite{Simplified_qutrit_circuit, Gokhale2019}, and implementation of ternary Shor’s \cite{Ternary_Shors, Ternary_Shors_2} and Grover's search algorithms \cite{Wang2011} with a more compact quantum circuit. Qudits have also been shown numerically and experimentally to offer increased resilience to decoherence \cite{Liu2009, Cozzolino2019,Ecker2019, tilly2021qudits}.
Experimentally, several successful implementations of qutrits have been demonstrated on ion traps \cite{Bruzewicz2019Trapped-ionChallenges, Ringbauer2022AIons}, photonic systems \cite{Fickler2012QuantumMomenta, Malik2016Multi-photonDimensions, Wang2018MultidimensionalOptics, Lu2020QuantumPhoton, Kues2017On-chipControl}, and superconducting circuits \cite{Peterer2015CoherenceQubit, Bianchetti2010ControlAtom, Dong2022SimulationQudit, Tan2018TopologicalQutrit}, with applications in qudit quantum algorithms and simulations \cite{Kues2017, Single_photon_qutrit_PEA, Blok2021,  Tan2021ExperimentalQudit, Cao2022EmulatingAlgorithms}. 
Theoretically, the use of qudits for QML has been suggested for quantum classifiers \cite{Quantum_classifier}, deep quantum neural networks \cite{DQNN}, and quantum reinforcement learning protocols \cite{Quantum_reinforcement_learning}. 

In this work we explore the potential of qutrit systems for classification tasks, and demonstrate a hardware implementation using a single superconducting qutrit. We propose a generalization of the NAE encoding to qudits and introduce two new encoding schemes. We propose an optimized encoding method, which integrates training into the encoding process by using a parameterized circuit to encode the data and subsequently applying a separate parameterized circuit for data classification. We perform ternary classification tasks on the well-studied Iris dataset \cite{iris}, where the obtained performance can be compared to previous results for qubit classifiers \cite{Quantum_classifier, Iris_multiclass} performing binary classification.
To exhibit the versatility and effectiveness of our qutrit classifiers beyond the widely used Iris dataset, we also benchmark our model using the Palmerpenguins dataset \cite{penguins}.
Through experiments on superconducting transmon devices we demonstrate 94.5\% ternary classification accuracy for the Iris dataset using a single qutrit.

\section{methods} \label{sec:methods}

\subsection{Qudit data encoding schemes}
\label{sec:MethodsEncoding}
To achieve classification using a parameterized quantum circuit, rescaled classical data is encoded into the quantum states, hence allowing the calculation of a loss function based on the results of measuring a chosen subset of qudits. 
The encoding scheme determines how the classical features get represented on a quantum computer, and is hence a critical component of designing a QML circuit. In this work, we propose a generalization to general qudits with arbitrary dimension $d$ of the qubit NAE scheme (referred to as qubit encoding in Ref. \cite{QCEncoding}). On one qudit it is expressed as
\begin{align}
\ket{\psi_\mathrm{NAE}}& = 
 \sum_{j=0}^{d-2} \left(\prod_{k=0}^{j-1}\sin{x_k}\right)\cos{x_j}\ket{j} 
    + \left(\prod_{k=0}^{d-2}\sin{x_k}\right)\ket{d-1} 
    , \label{eq_NAE}
\end{align}
where $x_{j(k)}$ denotes the $j(k)$-th feature in a rescaled classical data vector $\mathbf{x}$ of length $K$. On one qudit at most $d-1$ features can be encoded. For $K>d-1$, the encoding given in the above equation is applied on $N$ qudits to encode all entries in $\mathbf{x}$, where $N=\lceil K/(d-1) \rceil$ is given by rounding up $K/(d-1)$ to the nearest integer.

We introduce two new encoding schemes, non-entangled phase encoding (NPE) 
\begin{align}
\ket{\psi_\mathrm{NPE}} = \frac{1}{\sqrt{d}}\left(\ket{0}+\sum_{j=1}^{d-1}e^{ix_{j-1}}\ket{j}\right), \label{eq_NPE}
\end{align}
and non-entangled compact encoding (NCE)
\begin{align}
    \ket{\psi_\mathrm{NCE}} &=e^{ix_{2d-3}}\left(\prod_{k=0}^{d-2}\sin{x_k}\right)\ket{d-1}\nonumber\\
    +&\sum_{j=1}^{d-2} \left(e^{ix_{d-2+j}}\prod_{k=0}^{j-1}\sin{x_k}\right)\cos{x_j}\ket{j} + \cos{x_0}\ket{0}. \label{eq_NCE} 
\end{align}
NPE can encode $d-1$ features for each qudit, in analogy to NAE, and therefore the number of required qudits is the same. NCE is a dense encoding that combines the two, which can encode $2(d-1)$ features per qudit, so that the required number of qudits for NCE is given by $N=\lceil K/(2(d-1)) \rceil$. This is the maximum possible encoding density for a non-entangled qudit encoding. Importantly, for all encoding schemes the required number of qudits decreases with increasing $d$, which is one of the main potential advantages of using higher level qudits over qubits.

All the above encoding schemes can be implemented as $\ket{\psi} = U_e(\mathbf{x})\ket{0}$, where $U_e$ is a unitary operator describing the operation of a quantum circuit on $N$ qudits, and $\ket{\psi}$ denotes a general resulting encoded $N$-qudit state vector. We present the resulting specific equations and circuits for qutrits ($d=3$) in Appendix \ref{AppendixSubspaceGates}, and the quantum circuits for the general qudit encoding schemes in Appendix \ref{sesc:SMdataencodings}. 

The classical data is typically rescaled into the range $[0, \frac{\pi}{2}]$ for qubit encodings \cite{Quantum_classifier, Iris_multiclass}. For NAE (Eq. \ref{eq_NAE}), if one of the $x_{k}$ exactly rescales to 0, which is typically the case for the smallest value among each feature, then all following state amplitudes on that qudit containing $\sin{x_k}$ are 0, so that any information about $x_{l} \:\forall l>k$ is lost on that qudit.
Therefore, to avoid this loss of information we shift the rescaling range to be $[\frac{\pi}{4}, \frac{3\pi}{4}]$. We use the same rescaling for NCE and for consistency, although NPE does not have the problem of encoding information loss, we also shift its rescaling range accordingly.

\subsection{Optimized encoding}
\label{method_encoding_training}
\begin{figure}[htpb]
    \centering
    \begin{tikzpicture}
	\begin{pgfonlayer}{nodelayer}
		\node [style=none] (0) at (-7.5, 1) {$|0\rangle^{\otimes N}$};
		\node [style=none] (5) at (-5.5, 2) {};
		\node [style=none] (6) at (-5.5, 0) {};
		\node [style=none] (7) at (-1, 0) {};
		\node [style=none] (8) at (-1, 2) {};
		\node [style=none] (9) at (-3, -0.5) {Encoding circuit};
		\node [style=none] (10) at (2.25, -0.5) {Classifier circuit};
		\node [style=none] (15) at (0, 2) {};
		\node [style=none] (16) at (0, 0) {};
		\node [style=none] (17) at (4.5, 0) {};
		\node [style=none] (18) at (4.5, 2) {};
		\node [style=meter] (23) at (6, 1) {};
		\node [style=none] (24) at (-6.5, 1) {};
		\node [style=none] (25) at (-7.5, 5) {$|0\rangle^{\otimes N}$};
		\node [style=none] (30) at (-5.5, 6) {};
		\node [style=none] (31) at (-5.5, 4) {};
		\node [style=none] (32) at (-1, 4) {};
		\node [style=none] (33) at (-1, 6) {};
		\node [style=none] (34) at (-3.25, 3.5) {State preparation};
		\node [style=none] (35) at (-6.5, 5) {};
		\node [style=none] (40) at (0, 6) {};
		\node [style=none] (41) at (0, 4) {};
		\node [style=none] (42) at (4.5, 4) {};
		\node [style=none] (43) at (4.5, 6) {};
		\node [style=none] (44) at (2.25, 3.5) {State inversion};
		\node [style=meter] (45) at (6, 5) {};
		\node [style=none] (46) at (-9, 5) {(a)};
		\node [style=none] (47) at (-9, 1) {(b)};
		\node [style=none] (48) at (-3, 5) {};
		\node [style=none] (49) at (-3.25, 5) {$U_e(\boldsymbol{\phi})$};
		\node [style=none] (51) at (2.75, 5) {};
		\node [style=none] (52) at (2.25, 5) {$U_e^\dagger(\boldsymbol{\phi}^\prime)$};
		\node [style=none] (55) at (-3, 1) {};
		\node [style=none] (56) at (-3.25, 1) {$U_e(\boldsymbol{\phi})$};
		\node [style=none] (58) at (-5.5, 5) {};
		\node [style=none] (59) at (-1, 5) {};
		\node [style=none] (60) at (0, 5) {};
		\node [style=none] (61) at (4.5, 5) {};
		\node [style=none] (62) at (4.5, 1) {};
		\node [style=none] (63) at (0, 1) {};
		\node [style=none] (64) at (-1, 1) {};
		\node [style=none] (65) at (-5.5, 1) {};
		\node [style=none] (66) at (2.25, 1) {$U_c(\boldsymbol{\theta})$};
	\end{pgfonlayer}
	\begin{pgfonlayer}{edgelayer}
		\draw (30.center) to (31.center);
		\draw (31.center) to (32.center);
		\draw (32.center) to (33.center);
		\draw (33.center) to (30.center);
		\draw (40.center) to (41.center);
		\draw (41.center) to (42.center);
		\draw (42.center) to (43.center);
		\draw (43.center) to (40.center);
		\draw (5.center) to (6.center);
		\draw (6.center) to (7.center);
		\draw (7.center) to (8.center);
		\draw (8.center) to (5.center);
		\draw (15.center) to (16.center);
		\draw (16.center) to (17.center);
		\draw (17.center) to (18.center);
		\draw (18.center) to (15.center);
		\draw (35.center) to (58.center);
		\draw (59.center) to (60.center);
		\draw (61.center) to (45);
		\draw (24.center) to (65.center);
		\draw (64.center) to (63.center);
		\draw (62.center) to (23);
	\end{pgfonlayer}
\end{tikzpicture} 
    \caption{(a) Quantum circuit measuring overlap between two states: the probability of measuring the zero state on all qudits after application of the quantum circuit gives the value of $|\bra{0}U_e^\dagger(\boldsymbol{\phi}^\prime)U_e(\boldsymbol{\phi})\ket{0}|^2$. (b) Quantum circuit implementing classification: after encoding of the data with $U_e$, the classification unitary $U_c$ is implemented with parametrized quantum circuits, with parameters denoted as \boldsymbol{$\theta$}. }
    \label{fig:encoding_training}

\end{figure}
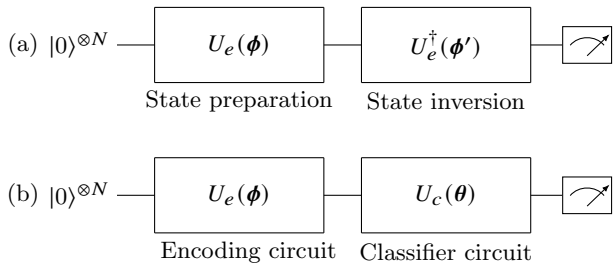
The features in the vector space often exhibit some geometric relations between each other, which contain information for each sample data point that can help in the classification task \cite{Wiebe2020}. While the encoding schemes (Eqs. \ref{eq_NAE}-\ref{eq_NCE}) map classical data to quantum states in an efficient way, they do not usually preserve such relations in the Hilbert space. Furthermore, when mapping classical features to quantum states within a given encoding scheme (Eqs. \ref{eq_NAE}-\ref{eq_NCE}), one has to choose which feature to assign to each integer index $k$.
One inherent limitation of this approach is that, when developing a QML algorithm, one does not have the a priori knowledge of the best feature assignment ordering. 
We therefore propose an additional training step to optimize the encoding to overcome this problem, where a new $K$-dimensional vector $\boldsymbol{\phi}$ is introduced to replace the basic feature vector $\mathbf{x}$ as $\boldsymbol{\phi} = \mathbf{Wx}+\mathbf{b}$,
where $\mathbf{W}$ is a $K$ by $K$ real matrix, and $\mathbf{b}$ is a $K$-dimensional real vector, so that $x_{j(k)}$ in Eqs. \ref{eq_NAE}-\ref{eq_NCE} is replaced by the corresponding $\phi_{j(k)}$.
As shown in the context of quantum state discrimination, the maximally achievable classification accuracy is obtained when the encoded states of different classes have the smallest possible overlap (or more generally, fidelity)\cite{QSDAndrew2020}.
Therefore, we construct the optimal $\mathbf{W}$ and $\mathbf{b}$ by minimizing a quantity related to the fidelity between states of different classes. While one could in principle minimize the fidelity itself, the drawback would be that it is very time-consuming to compute, we therefore minimize a related quantity which can be computed more efficiently.

Each class in the classification task has an associated set of data points, which we denote as $\mathbf{\mathcal{X}}_i$, with $i$ being the class label. If we randomly sample $m_i$ data points from each set $\mathbf{\mathcal{X}}_i$, the collection of the encoded states for these samples can be considered a mixed state, and hence represented as a density matrix 
\begin{align}
    \rho_i = \frac{1}{m_i}\sum_{n=1}^{m_i} \ket{\psi_{i,n}}\bra{\psi_{i,n}},
\end{align}
where $\ket{\psi_{i,n}}= U_e(\boldsymbol{\phi}_{i,n})\ket{0}$ is the encoded state produced with an encoding circuit $U_e$ acting on $\boldsymbol{\phi}_{i,n} = \mathbf{W x}_{i,n}+\mathbf{b}$, and $\mathbf{x}_{i,n} \in \mathbf{\mathcal{X}}_{i}$.

We then find $\mathbf{W}$ and $\mathbf{b}$ by minimizing the loss function
\begin{align}
    \mathcal{L}_{e} =  \sum_{i}\sum_{j\neq i}\mathrm{Tr}\left[\rho_i \rho_j\right]^2 -  \sum_{i}\mathrm{Tr}\left[\rho_i \rho_i\right]^2.
\label{eq_loss_encoding_training}
\end{align}
The first term of $\mathcal{L}_{e}$ minimizes the quantity related to the overlap between states of two different classes. If both $\rho_i$ and $\rho_j$ are pure states, $\mathrm{Tr}\left[\rho_i \rho_j\right]$ represents the squared overlap between these two states. The second term maximizes the purity of the states in each class, where the purity of $\rho_i$ is defined as $\mathrm{Tr}[\rho_i \rho_i]$. The calculation of $\mathrm{Tr}\left[\rho_i \rho_j\right]$ involves averaging the overlap between each state pair $\bra{\psi_{j,n}}$ and $\ket{\psi_{i,n^\prime}}$ in classes $i$ and $j$. 
Note that although the quantity $\mathrm{Tr}\left[\rho_i \rho_j\right]$ represents a property similar to the fidelity, it is not an exact measure of the fidelity between mixed states $\rho_i$ and $\rho_j$. We employ this quantity due to its lower computational cost on quantum processors, since calculating the fidelity itself would be prohibitively time-consuming.
We further square the values in the loss function to impose greater penalties on high overlaps and low purities.

The value of $\mathrm{Tr}\left[\rho_i \rho_j\right]$ can be obtained in experiments on quantum hardware by expressing it as
\begin{align}
     \mathrm{Tr}\left[\rho_i \rho_j\right]= \frac{1}{m_i m_j} \sum_{n,n'} | \bra{0}U_e^\dagger(\boldsymbol{\phi}_{i,n})U_e(\boldsymbol{\phi}_{j,n'})\ket{0} |^2.
     \label{eq:Trrhoirhoj}
\end{align}
Each of the terms in the sum can be measured on a quantum computer by executing two subsequent encoding circuits as shown in Fig. \ref{fig:encoding_training}a, and then measuring the probability to obtain the $\ket{0}$ state. Note that when scaled to a larger number of qudits and/or a larger dataset, the number of expectation values required to estimate $\mathrm{Tr}\left[\rho_i \rho_j\right]$ does not scale with the number of qudits; it only scales with respect to the number of samples $m_i$ and $m_j$.

Once the optimal values of $\mathbf{W}$ and $\mathbf{b}$ are found by minimizing $\mathcal{L}_{e}$ (Eq. \ref{eq_loss_encoding_training}), they are fixed for the subsequent optimization of the classifier part of the circuit (Fig. \ref{fig:encoding_training}b). 

\subsection{Classifier optimization}
The optimal quantum circuit parameters are found by minimizing a task-specific loss function. By repeating the process of applying the trained quantum circuit to an encoded data point and then  measuring a subset of qudits, the predicted class label is inferred based on the most likely measurement outcome. 

For our numerical simulations we use a squared error loss function, given by
\begin{align}
    \mathcal{L}_{SE}=\sum_i (1-P(\ket{y_i}))^2,
    \label{eq_LSE}
\end{align}
where the integer $y_i\in \{0, 1, 2\}$ is the class label of the $i$-th data point, and $P(\ket{y_i})$ denotes the probability of measuring the state corresponding to $y_i$. The special case of the single qubit loss function is discussed in Appendix \ref{AppxOptimization}. In our hardware implementation we use a linear loss function, given by
\begin{align}
    \mathcal{L}_c = \sum_i 1-P(\ket{y_i}).
    \label{eq:lossLinear}
\end{align}
This is due to the fact that in the hardware runs we use a rotosolve-like method for the parameter optimization, which requires the cost function of every single variable to be a sinusoidal function \cite{Nakanishi2020Sequential,Ostaszewski2021structure}; this is the case for the linear loss function, but not for the squared loss function. Note that both the square and linear loss functions give the same minimum.

\section{Numerical simulation of the qutrit classifier}
\label{sec:results}
\subsection{Datasets}
Our target datasets for ternary classification are the Iris dataset \cite{iris} and the Palmerpenguins dataset \cite{penguins}. Both datasets comprise four numerical features, pertaining to Iris flowers for the Iris dataset and to penguins for the Palmerspenguins dataset, with data distributed across three classes of species. The Iris dataset, widely studied in classical machine learning literature, has previously served as a target dataset for binary classification using qubit classifiers \cite{Quantum_classifier, Iris_multiclass}. The additional use of the Palmerpenguins dataset allows us to demonstrate the adaptability and efficiency of our qutrit classifiers.

\subsection{Encoding and Classifier circuits}
\label{sec:QutritCircuits}
To encode the four features, for NAE and NPE the number of qudits is $N=\lceil 4/(d-1) \rceil$, and for NCE it is $N=\lceil 2/(d-1) \rceil$ (see Sec. \ref{sec:MethodsEncoding}). We therefore need four qubits ($d=2$) when using NAE or NPE, and two qubits when using NCE. For qutrits ($d=3$) the required numbers are halved to two for NAE and NPE, and to only one for NCE. In the physical realization of superconducting circuit processors, both qutrits and qubits correspond to identical transmon physical hardware components within the system, and they differ in the number of states that are accessed. A general advantage of operating the transmons as qutrits therefore is that less transmons are required than when they are operated as qubits. For the considered datasets, which require encoding four features, one transmon suffices for qutrit operation, while two are needed when operated as qubits. We achieve this by using our dense NCE encoding for qutrits and qubits. In literature the NAE has been used to perform binary classification on the Iris dataset, which required four qubits, and hence double what we require when using NCE \cite{Quantum_classifier}.

To determine how the type of encoding affects the qutrit classification, we also consider NAE and NPE encoding on two qutrits for a systematic comparison to the one qutrit NCE results. 
We further aim to compare the classification accuracy of qubits and qutrits when the same number of these physical building elements is used. To perform this comparison for both one and two transmons, we perform also NCE encoded classification for only one qubit, where only two of the four features can be encoded. This allows us to evaluate how such a loss of information as part of the encoding into only two instead of three levels affects the classification accuracy. In summary, we therefore perform single-qubit, single-qutrit, and two-qubit classification using NCE, and two-qutrit classification using NAE and NPE. 

\begin{figure}[htpb]
    \centering
    \includegraphics[width=0.49 \textwidth]{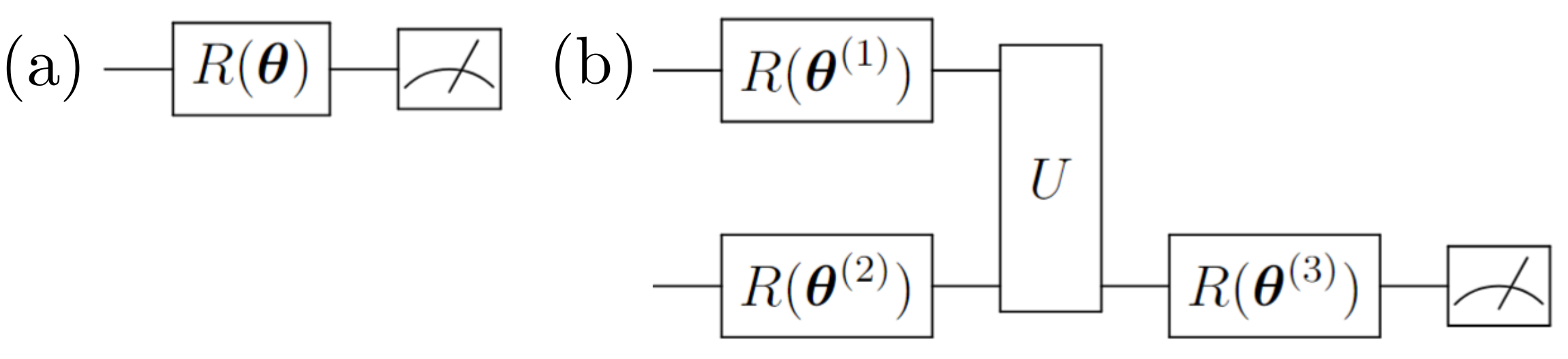}
    \caption{General TTN structures for (a) one-qudit and (b) two-qudit classifiers, where $R(\boldsymbol{\theta})$ gates represent general single-qudit unitaries, and $U$ represents an arbitrary two-qudit unitary. }
    \label{f_general_qudit_classifier}
\end{figure}

\begin{figure}[htpb]
    \centering
    \includegraphics[width=0.47 \textwidth]{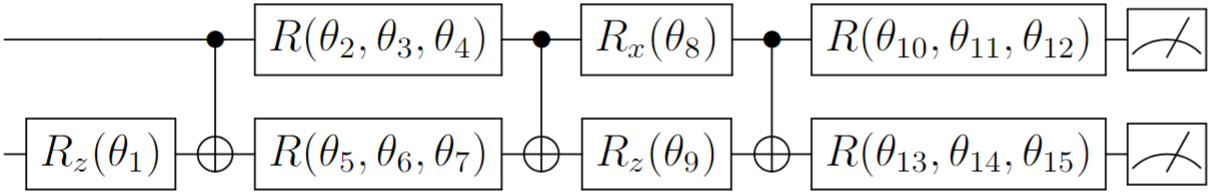}
    \caption{Circuit implementing a general two-qubit classifier, where $R(\theta_1,\theta_2,\theta_3)=R_z(\theta_3)R_x(\theta_2)R_z(\theta_1)$.}
    \label{f_general_two_qubit}
\end{figure}

Our proposed parametric qutrit classifier circuits follow the tree-like tensor network (TTN) structure \cite{TTN}, which has been previously used as an ansatz for qubit ML circuits \cite{Quantum_classifier}. In TTN structures, general parametrized single-qudit operations are combined with two-qudit gates to progressively halve the number of operational qudits. 
The general TTN structures for one-qudit and two-qudit classifiers are shown in Fig. \ref{f_general_qudit_classifier}, where $R(\boldsymbol{\theta})$ gates represent general single-qudit unitaries, and $U$ represents an arbitrary two-qudit unitary. The length of the vector $\boldsymbol{\theta}$ is at least $d^2-1$ to enable the construction of a general $\text{SU}(d)$ unitary. For qubits we choose $R(\theta_1,\theta_2,\theta_3)=R_z(\theta_3)R_x(\theta_2)R_z(\theta_1)$, and for the two-qubit gate $U$ we choose the CNOT gate. 
The two-qubit ansatz derived from the TTN structure in Fig. \ref{f_general_qudit_classifier}b therefore has 9 parameters, fewer than the 15 needed for a universal two-qubit operation \cite{shende_minimal_2004}. Hence we additionally compute the two-qubit classification using the universal two-qubit circuit proposed in Ref. \cite{shende_minimal_2004}, and shown in Fig. \ref{f_general_two_qubit}. This circuit has the largest possible expressivity on two qubits, and hence allows to obtain the largest classification accuracy possible on two qubits.

There are many equivalent representations of a general single qutrit unitary, such as the one based on the Gell-Mann matrix representation \cite{gellmann}. Using gates that can be implemented on hardware for superconducting qutrits \cite{Blok2021} (see Appendix \ref{AppendixSubspaceGates} for detailed descriptions), a general single qutrit operation $R_L(\boldsymbol{\theta}_{1\cdots8})$ can be constructed as \cite{Clements2016}
\begin{align}
    R_L(\boldsymbol{\theta}_{1\cdots8}) = & R_{z01}(\theta_8) \cdot R_{x01}(\theta_7) \cdot R_{z12}(\theta_6) \cdot R_{x12}(\theta_5) \cdot  \nonumber \\ & R_{z12}(\theta_4) \cdot R_{z01}(\theta_3) \cdot R_{x01}(\theta_2) \cdot R_{z01}(\theta_1),
\label{eq_RL_theoretical}
\end{align}
where $R_{zuv}$ and $R_{xuv}$ are $Z$ and $X$ rotations in the two-level subspace $\{\ket{u}, \ket{v}\}$ of the qutrit (see Appendix \ref{Appx_decomposition} for the steps of decomposing an arbitrary SU(3) unitary into $R_L$). 
For increased efficiency of the hardware runs on our superconducting transmons we use slightly modified equivalent forms of the $Z$ and $X$ rotations, described in what follows.
Similarly to a physical single qubit gate, our physical single qutrit gate is implemented by driving a microwave pulse that transfers a state between the $\{\ket{0},\ket{1}\}$ set of states or between $\{\ket{1},\ket{2}\}$, corresponding to matrix exponentials of Gell-Mann matrices.
As an extension of the qubit virtual $Z$ gate~\cite{McKay2017EfficientZGates}, our qutrit virtual $Z$ gate is a shift of the phases of all following pulses. An additional constraint for the choice of hardware gates is the fact that in the hardware runs we use a rotosolve-like optimization method, where every gate needs to have exactly two different eigenvalues~\cite{Nakanishi2020Sequential,Ostaszewski2021structure}. To ensure this, we define following gates as the basic building blocks: $R_{y01}(\theta)=e^{-i \frac{\theta}{2} \lambda_2}$, $R_{y12}(\theta)=e^{-i \frac{\theta}{2} \lambda_7}$, $H_1= Z_1(\pi)R_{y01}(-\pi/2)$, $H_2= Z_2(\pi)R_{y12}(-\pi/2)$, $X'_{01}(\theta)= H_1 Z_1(\theta) H_1$, $X'_{12}(\theta)= H_2 Z_2(\theta) H_2$,  where $\lambda_2$ and $\lambda_7$ are Gell-Mann matrices, $H_1$ and $H_2$ are two-level Hadamard gates on the 0 and 1 levels or on the 1 and 2 levels. We add the prime subscript on the $X'_{uv}$ gates to emphasize their difference from the basic $R_{xuv}$ rotations. With these building blocks, we implement the general $R_L$ operation on hardware as 
\begin{align}
    R_L'(\boldsymbol{\theta}_{1\cdots8}) = & X'_{01}(\theta_8) \cdot Z_{1}(\theta_7) \cdot  Z_{2}(\theta_6) \cdot X'_{12}(\theta_5) \cdot \nonumber \\ &  Z_{2}(\theta_4) \cdot Z_{1}(\theta_3) \cdot X'_{01}(\theta_2) \cdot Z_{1}(\theta_1).
\label{eq_RL_hardware}
\end{align}
In Appendix \ref{Appx_equivalence} we show the equivalence between this representation and the one presented in Eq. \ref{eq_RL_theoretical}.

As two-qutrit unitary we choose the SUM gate, which is a generalization of the CNOT gate to qutrits, and is defined as $\text{SUM}(\ket{a,b}) = \ket{a, (a+b)\text{ mod 3}}$.

\subsection{Classification results}
\begin{figure}[htpb]
    \centering
    \includegraphics[width=0.49 \textwidth]{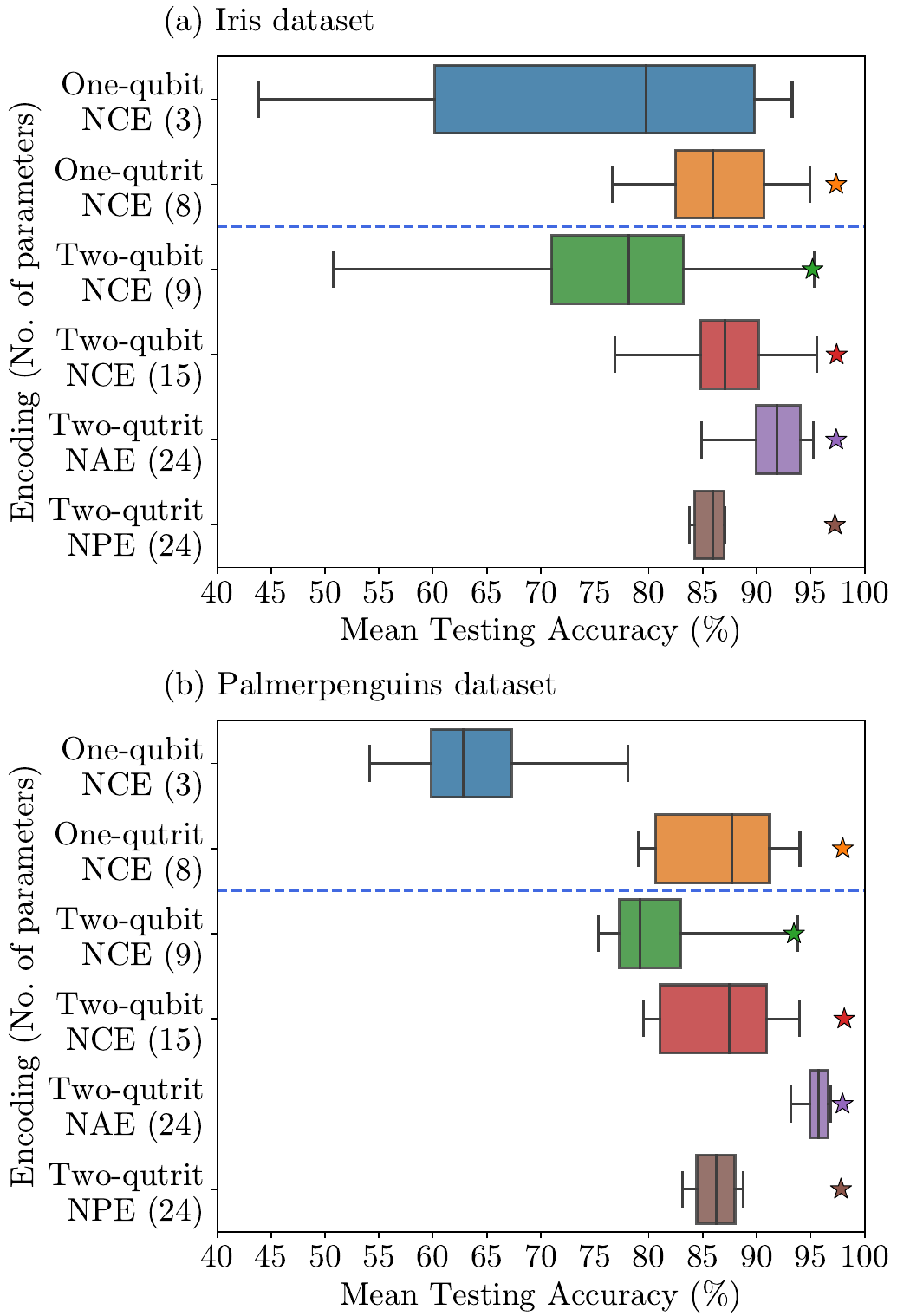}
    \caption{Box plots of the distributions of mean testing accuracy for different numbers of devices and encodings, for both the (a) Iris and (b) Palmerpenguins datasets. Each bar shows the minimum, the lower quartile, the median, the upper quartile, and the maximum of the mean testing accuracy of a given category, with a box drawn from the lower to the upper quartile. We compare one-qubit, one-qutrit, two-qubit and two-qutrit circuits for different encodings, with the number of parameters shown in brackets. For two-qubit circuits we compare a short ansatz derived from the TTN structure with 9 parameters (Fig. \ref{f_general_qudit_classifier}b) and a long ansatz that constructs a universal two-qubit operation with 15 parameters (Fig. \ref{f_general_two_qubit}). The stars show the accuracy obtained using our optimized encoding approach, all of which have a standard deviation within 2\%, which is within the extension of the stars in the graph.}
    \label{f_main}
\end{figure}
As discussed in the first part of the previous subsection, we perform single-qubit, single-qutrit, and two-qubit classification using NCE, and two-qutrit classification using NAE and NPE. This allows us to perform a systematic analysis of how classification accuracy is influenced by three factors: the type of qudit (qubit or qutrit), the number of qudits (one or two), and the chosen encoding (NAE, NPE, NCE).
Furthermore, to evaluate the effect of the expressibility of the classifier circuit, for two-qubit classification we use the short TTN-based ansatz with 9 parameters (Fig. \ref{f_general_qudit_classifier}b) and the longer universal two-qubit circuit (Fig. \ref{f_general_two_qubit}).

The order of the individual features within the feature vector $\mathbf{x}$ can be chosen arbitrarily (see Sec. \ref{method_encoding_training}), and may significantly affect the maximally achievable classification accuracy. To evaluate the range of changes in classification accuracy for different orderings, for all qudit encodings we perform separate classification for all of the possible orderings in which data features are permuted when assigned to the encoding gates. There are 4 features, resulting in 4!=24 permutations. Note that for special case of the single qubit results, where even using NCE only two features can be encoded, there are only 12 possibilities when choosing two features out of four and permuting them. For each permutation we train the classification circuit and calculate the average testing accuracy for 50 random splits of training and testing data with a ratio of 2:1. 
Details of the optimization process are presented in Appendix \ref{AppxOptimization}. In Fig. \ref{f_main} we show the resulting classification accuracy distributions in the form of box plots for all the permutations within each qudit encoding, both for the Iris and for the Palmerpenguins datasets.

The first thing we note is that the ordering of the data features as part of the encoding leads to a very large spread of obtained accuracies, which also varies significantly for different encoding methods. Single qubit results have an additional spread due to the fact that only two features can be encoded, and the accuracy varies significantly depending on the choice of the two features. Importantly, for both datasets, the one-qutrit classifier consistently outperforms the one-qubit classifier in all metrics, including the minimum, median and maximum mean testing accuracy among all permutations, as well as having a smaller spread of accuracy across permutations. The performance difference is due to the one-qubit classifier not being able to encode all four features of the data, whereas the one-qutrit classifier with NCE can. This confirms that operating a transmon as qutrit instead of as qubit can lead to significantly increased classification accuracy.

When using two qubits or qutrits, the classification accuracy is in a similar range as the one of the one-qutrit classifier. The two-qubit classifier with the shorter circuit (green bar in Fig. \ref{f_main}) has a somewhat lower accuracy, which is due to the reduced expressibility of the classifier part of its circuit. For two qutrits, NAE outperforms NPE, while the performance of the two-qubit classifier with longer circuit depth is between these two. The large dependence of the classification accuracy on the encoding, and on the data ordering permutations for a given encoding, raises the question as to whether an even larger classification accuracy could be obtained by other types of encodings and orderings. This uncertainty prevents the determination of whether for two transmons the qutrit performance is better than or equal to qubit performance, and whether the two-qutrits performance is higher than or equal to the one-qutrit performance. To overcome this problem an encoding method that systematically gives high accuracy levels with small spread is needed. In the next subsection we will show that our optimized encoding scheme achieves this.

\subsection{Training the optimal encoding}
We apply our optimized encoding scheme presented in Sec. \ref{method_encoding_training} to this system, and evaluate if it can systematically give a high classification accuracy. Note that we do not not apply the optimized encoding scheme to one-qubit classifiers, since for the ternary classification task considered here the optimized encoding scheme performs a minimization of the overlap between states of three classes, while on a single qubit only at most two classes of states can be orthogonalized to each other. 

The Iris and the Palmerpenguins datasets have 150 and 333 data points, respectively. We take 2/3 of the data points as the training set for the encoding optimization, and the remaining data as testing set. Since the number of data points is small, we use all points in the training set to calculate the mixed state density matrix representing encoded states in each class, instead of randomly sampling from the training set. After the encoding is trained, the optimization of the classification circuit is carried out in the same way as in the previous subsection.
Each setup is benchmarked with 50 random splittings into training and testing data sets, and the mean testing accuracy is calculated (see Appendix \ref{AppxOptimization}).

We plot the result as stars in Fig. \ref{f_main}. The standard deviation for the 50 random initializations with the optimized encoding is always within 2\%, and is therefore within the extension of the stars in the graph. The optimized encoding consistently improves the classification performance, and the achieved accuracy achieved either surpasses or matches the largest value obtained with the fixed encoding schemes. Importantly, our results show that when using the optimized encoding, the choice of underlying encoding scheme does not impact the classification performance: all classifiers, except the short two-qubit circuit, result in an accuracy close to 97\%. Furthermore, the small variation in the results indicates the robustness of the optimal encoding to the choice of the underlying encoding scheme. The reduced accuracy of the short two-qubit classifier can be attributed to its limited expressibility of the classifier part of the circuit. 

With the optimized encoding, the one-qutrit classifier achieves a similar accuracy as the two-qubit and two-qutrit classifiers. This finding further indicates that fewer qudits with a higher $d$ value can efficiently replace a larger number of qudits with a lower $d$, provided the classifier part of the circuit maintains adequate expressiveness.

\begin{figure*}[hbtp]
    \centering
        \hfill
        \includegraphics[width=0.45\textwidth]{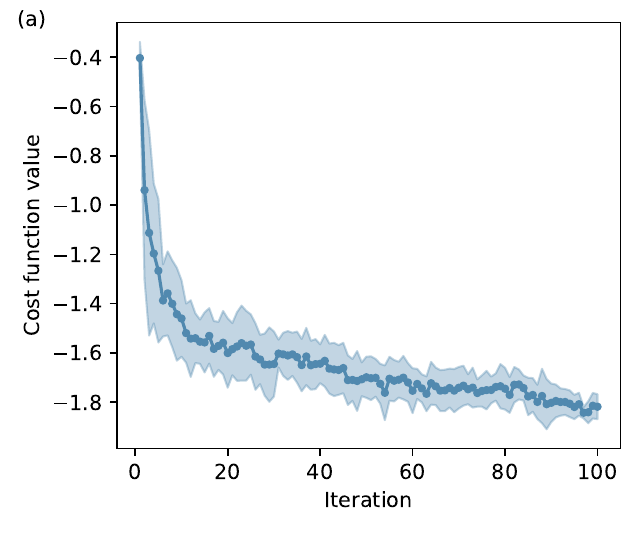}          
        \hfill
        \includegraphics[width=0.53\textwidth]{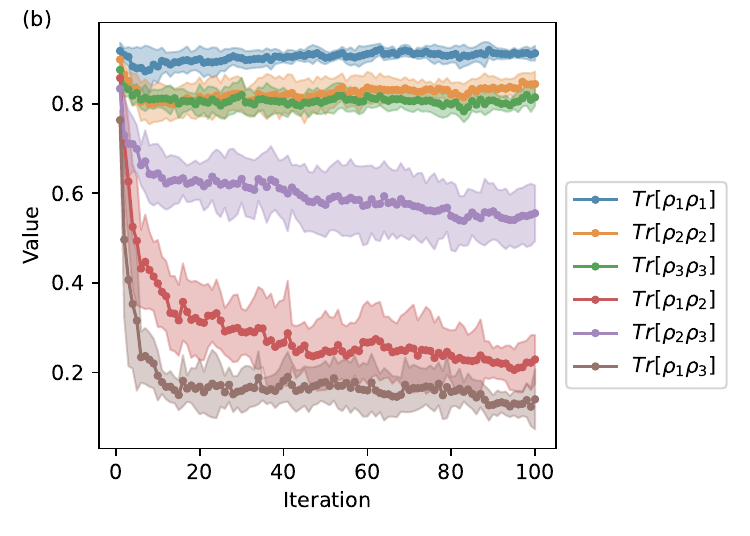}
        \hfill
    \caption{Performance of the encoding training process: the shadow denotes the standard deviation over 8 runs with different random seeds to denote the error bar. (a) Cost function value (Eq. \ref{eq_loss_encoding_training}) evaluated versus training iterations.  (b) Estimated class purity $\mathrm{Tr}[\rho_i\rho_i]$ and overlaps $\mathrm{Tr}[\rho_i\rho_{j\neq i}]$ versus training iterations. After 100 training steps the values of $\mathrm{Tr}[\rho_1\rho_1]$,$\mathrm{Tr}[\rho_2\rho_2]$ and $\mathrm{Tr}[\rho_3\rho_3]$ are $0.91 \pm 0.02 $ ,$0.84 \pm 0.03 $, $0.81 \pm 0.02 $, respectively. The values of $\mathrm{Tr}[\rho_1\rho_2]$,$\mathrm{Tr}[\rho_2\rho_3]$,$\mathrm{Tr}[\rho_1\rho_3]$ are $0.23\pm 0.06 $, $0.56\pm 0.07 $, $0.14\pm 0.08 $, respectively.}
    \label{fig:encoding_training_}
\end{figure*}

\begin{figure}[hbt!]
    \includegraphics[width=\linewidth]{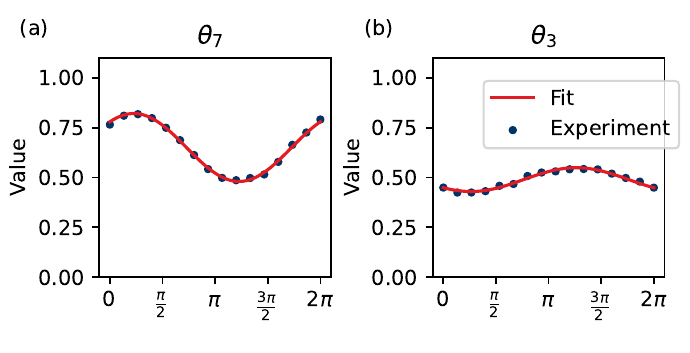} \\

    \includegraphics[width=\linewidth]{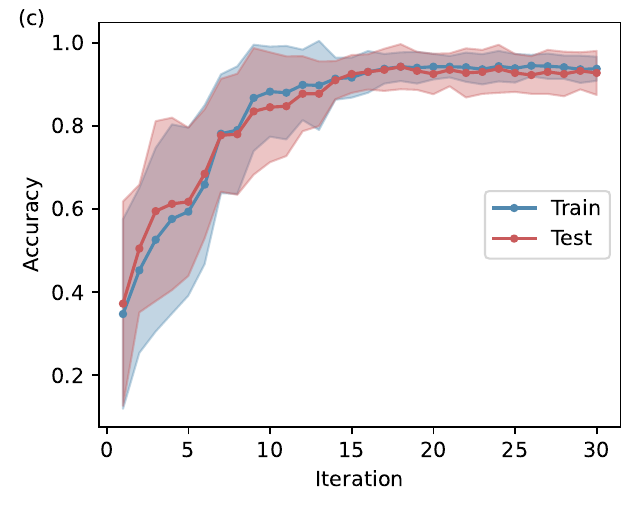}
    \caption{(a-b) A typical rotosolve-like optimization iteration during the training process: the optimization algorithm sweeps each parameter from 0 to $2\pi$ and fits the result with a sinusoidal function; the parameter value is updated to be the one at the minimum of the fitted function. (c) The classification accuracy for training and testing datasets versus rotosolve iteration number: the experiment is repeated 8 times with different random seeds; the solid line shows the averaged accuracy among all experiments, and the shadowed area shows the maximum and minimum accuracy across all experiments. The accuracy values for the training and testing sets after 32 iterations are $0.96\pm 0.03$ and $0.94\pm 0.04$, respectively.}
    \label{fig:rotosolve_accuracies}
\end{figure}

\section{Hardware implementation and results}
Given the good performance of the one-qutrit classifier with NCE in emulations when combined with optimized encoding, we evaluate whether it can be successfully implemented in currently available noisy hardware. To this aim we execute it to classify the Iris dataset on a single fixed-frequency superconducting transmon~\cite{Rahamim2017Double-sided,spring2021high}. The experimental setup and the device characterization have been reported in Refs. \cite{Cao2022EfficientTransmon,Cao2022EmulatingAlgorithms}. We use the $X'_{uv}$ and $Z_v$ based qutrit gate-set described in Sec. \ref{sec:QutritCircuits}, and the one-qutrit ansatz for encoding and classification is as shown in Fig. \ref{fig:encoding_training}, with $U_e(\boldsymbol{\phi})=Z_2(\phi_3)Z_1(\phi_1)X'_{12}(\phi_1)X'_{01}(\phi_0)$ and $U_c(\boldsymbol{\theta})=R'_L(\boldsymbol{\theta}_{1\cdots8})$.

We perform a total of eight distinct experiments, each using a different seed value to randomly select the 100 data points for the training set and the 50 data points for the testing set across all three classes. The different random seeds are also utilized to randomize the intrinsic parameters of the optimizer. The encoding optimization scheme iteratively determines the values of $\mathbf{W}$ and $\mathbf{b}$ by minimizing the loss function in Eq. \ref{eq_loss_encoding_training}. We randomly select 500 samples of pairs $x\in \mathcal{X}_i$ and $x^\prime \in \mathcal{X}_j$ in order to estimate the distribution of the overlap-like quantity in Eq. \ref{eq_loss_encoding_training}, $\mathrm{Tr}\left[\rho_i \rho_j\right]$. 
For given values of $\mathbf{W}$ and $\mathbf{b}$, 10 shots are executed for each pair of samples, and the overall measured probabilities of the qutrit being in the $\ket{0}$ state are used to estimate $\mathrm{Tr}\left[\rho_i \rho_j\right]$ using Eq. \ref{eq:Trrhoirhoj}.

The optimization of the loss function in Eq. \ref{eq_loss_encoding_training} is implemented with the Simultaneous Perturbation Stochastic Approximation (SPSA) method~\cite{Spall1998ANOO}. The SPSA algorithm is a method that approximates the gradient using a small random vector to perturb the objective function, and which needs to evaluate two points per iteration. The resulting experimental values for the loss function and for the quantities $\mathrm{Tr}\left[\rho_i \rho_j\right]$ ($i\ne j$) and $\mathrm{Tr}\left[\rho_i^2\right]$ versus the training steps are shown in Fig. \ref{fig:encoding_training_}. The experiment is repeated 8 times with different random seeds for training versus testing dataset splitting, parameter initialization, and random sampling pairs $\{x,x^\prime\}$. The solid points indicate the average value across all 8 experiments, and the boundary of the shadowed area denotes the maximum and minimum values. We terminate the training after 100 perturbation steps, at which stage the average purities of classes 1, 2, and 3 are $0.91\pm0.02$, $0.84\pm0.03$, and $0.81\pm0.02$, respectively. The average of the overlap-like quantity $\mathrm{Tr}\left[\rho_i \rho_j\right]$ between classes 1 and 2, 2 and 3, 1 and 3 are $0.23\pm0.06$, $0.56\pm0.07$, and $0.14\pm0.08$, respectively. These results are comparable to our results obtained in the numerical simulations for noiseless qutrits.

With the data encoding circuit trained and fixed, we proceed to train the classifier circuit using the rotosolve method~\cite{Ostaszewski2021structure,Nakanishi2020Sequential}. The parameters of the circuit are optimized to minimize the loss function $\mathcal{L}_c$ in Eq. \ref{eq:lossLinear}. The loss function value is estimated by taking the average over 10 shots (measurements) for each data point in the training dataset. During training, we modify the parameters of the classifier circuit one at a time, iteratively sweeping each parameter of the circuit while fixing the others. Each sweep samples 16 points from $-\pi$ to $\pi$, and fits the loss value with a sinusoidal function. Once fitted, we take the parameter that gives the minimum of the cost function as the selected value for the current parameter of the current iteration. A typical training result is shown in Fig. \ref{fig:rotosolve_accuracies}a and Fig. \ref{fig:rotosolve_accuracies}b.

After each iteration of fitting and updating all parameters to their current optimum values, we evaluate the classifier's performance with a larger shot number (500 shots). The resulting classification accuracy versus the number of rotosolve iterations is shown in Fig. \ref{fig:rotosolve_accuracies}c. The averaged accuracy starts at around $1/3$, which is the probability that a class gets its correct label with random assignment. Training stops after 32 rotosolve iterations since the reduction in the loss for more iterations becomes negligible. Each iteration comprises a complete sweep of one randomly selected parameter from the eight available. No parameter is chosen twice before all others are swept. Therefore, in 32 iterations, each parameter is swept four times. We have run the classifier circuit training 8 times, corresponding to using the same random seed for data splitting and the data encoding from the encoding training step. The solid points denote the average value, and the boundary of the shadow denotes the maximum and minimum values. The final average testing accuracy achieves 94.5\%$\pm$3.4\%. This result agrees well with the testing accuracy obtained on the qutrit emulator at 97.4\%$\pm$1.9\%, demonstrating that the method can be run on existing noisy hardware.

\section{Discussion}
In this study we present an implementation of a quantum machine learning classification task on a superconducting qutrit. We introduce three encoding schemes for qudits, and develop a optimized encoding scheme that can be efficiently implemented on quantum hardware. We show that only with such an optimized encoding one can consistently obtain high classification accuracy. Performing both numerical simulations and physical experiments, we show that using a reduced number of higher-dimensional physical qudits, such as qutrits, can effectively replace multiple qubits for quantum machine learning tasks. 

The answer to the question as to whether qutrits, or more generally higher-level system qudits, are more advantageous than qubits for quantum machine learning tasks as system sizes scale up, primarily depends on the extent to which known limitations for qubits, such as the appearance of vanishing gradients in the optimization process \cite{McClean2018}, are either exacerbated or mitigated when using higher-level qudits. Moreover, it is important to minimize experimental noise to take advantage of the lower number of qudits in a circuit compared to qubits. Our results demonstrate that noise levels in existing superconducting qutrits are low enough to allow for high classification accuracy on a single qutrit. It will be important to further investigate how this extends to circuits with a larger number of qutrits.

\acknowledgements
S.C. ~acknowledges support from the Eric and Wendy Schmidt AI in Science Postdoctoral Fellowship, a Schmidt Futures program. P.L.~acknowledges support from the EPSRC b. [EP/T001062/1, EP/N015118/1, EP/M013243/1]. M.B. acknowledges support from EPSRC QT Fellowship grant [EP/W027992/1]. I.R., A.A., and W. Z. acknowledge the support of the UK government Department for Science, Innovation and Technology through the UK National Quantum Technologies Programme.

\bibliography{bibli}

\appendix
\onecolumngrid

\section{Details of quantum gates and circuits\label{app:circuits}}
In this appendix we provide supplementary details of quantum gates and circuits used throughout the article, including the specific gate sets used in emulations and in the transmon experiment.

\subsection{Definitions of qubit and qutrit gates and encoding circuits}
\label{AppendixSubspaceGates}
In qubit classifiers we use the following gates: the X rotation $R_x(\theta)= \begin{pmatrix}
\cos{\frac{\theta}{2}} & -i\sin{\frac{\theta}{2}} \\ -i\sin{\frac{\theta}{2}} & \cos{\frac{\theta}{2}} \end{pmatrix}$, the Y rotation $R_y(\theta)= \begin{pmatrix}
\cos{\frac{\theta}{2}} & -\sin{\frac{\theta}{2}} \\ \sin{\frac{\theta}{2}} & \cos{\frac{\theta}{2}} \end{pmatrix}$, the Z rotation $R_z(\theta)= \begin{pmatrix} e^{-i\frac{\theta}{2}} & 0 \\ 0 & e^{i\frac{\theta}{2}}\end{pmatrix}$, the virtual phase shift $Z(\theta)= \left( \begin{array}{cccc}
1 & 0\\
0 & e^{i\theta}\end{array} \right)$, the Hadamard gate $H$ and the CNOT gate. In this article, for qubits we only use NCE (Eq. \ref{eq_NCE} with $d=2$), where on each qubit we can encode two features as follows:
\begin{flalign}
    \ket{\psi_\mathrm{NCE}}&= \cos{x_0}\ket{0}+e^{ix_1}\sin{x_0}\ket{1},&&
    \label{eq_d2NCE}
\end{flalign}
which can be implemented in a quantum ciruit as
\begin{flalign}
\nonumber & \:\:\: \Qcircuit @C=1.2em @R=1em {
\ket{0} & & \gate{R_{y}(2x_0)} & \gate{Z(x_1)} & \qw 
} &
\end{flalign}

In qutrit classifiers, we use the following single-qutrit rotations in the $\{\ket{0}, \ket{1}\}$ and $\{\ket{1}, \ket{2}\}$ subspaces, which are exponentials of Gell-Mann matrices $\lambda_{1\cdots8}$ \cite{gellmann}:

\vspace{0.5em}
\noindent $R_{x01}(\theta) = e^{-i \frac{\theta}{2} \lambda_1} = \begin{pmatrix}
\cos{\frac{\theta}{2}} & -i\sin{\frac{\theta}{2}} & 0 \\ -i\sin{\frac{\theta}{2}} & \cos{\frac{\theta}{2}} & 0 \\
0 & 0 & 1\end{pmatrix}$, $R_{x12}(\theta) = e^{-i \frac{\theta}{2} \lambda_6} = \begin{pmatrix}
1 & 0 & 0 \\
0 & \cos{\frac{\theta}{2}} & -i\sin{\frac{\theta}{2}} \\ 
0 & -i\sin{\frac{\theta}{2}} & \cos{\frac{\theta}{2}}
\end{pmatrix}$,

\vspace{0.5em}
\noindent $R_{y01}(\theta) = e^{-i \frac{\theta}{2} \lambda_2} = \begin{pmatrix}
\cos{\frac{\theta}{2}} & -\sin{\frac{\theta}{2}} & 0 \\
\sin{\frac{\theta}{2}} & \cos{\frac{\theta}{2}} & 0 \\ 
0 & 0 & 1
\end{pmatrix}$, $R_{y12}(\theta) = e^{-i \frac{\theta}{2} \lambda_7} = \begin{pmatrix}
1 & 0 & 0 \\
0 & \cos{\frac{\theta}{2}} & -\sin{\frac{\theta}{2}} \\ 
0 & \sin{\frac{\theta}{2}} & \cos{\frac{\theta}{2}}
\end{pmatrix}$, 

\vspace{0.5em}
\noindent $R_{z01}(\theta) = e^{-i \frac{\theta}{2} \lambda_3} = \begin{pmatrix}
e^{-i\frac{\theta}{2}} & 0 & 0 \\
0 & e^{i\frac{\theta}{2}} & 0 \\
0 & 0 & 1\end{pmatrix}$, $R_{z12}(\theta) = e^{-i \frac{\theta}{2} (\sqrt{3}\lambda_8-\lambda_3)/2} = \begin{pmatrix}
1 & 0 & 0 \\
0 & e^{-i\frac{\theta}{2}} & 0 \\ 
0 & 0 & e^{i\frac{\theta}{2}}\end{pmatrix}$,

\vspace{0.5em}
\noindent where the form of $R_{z12}(\theta)$ is chosen for consistency with $R_{z01}(\theta)$.\\

\vspace{0.5em}
\noindent We also use virtual phase shifts $Z_1(\theta)= \left( \begin{array}{cccc}
1 & 0 & 0\\
0 & e^{i\theta} & 0\\
0 & 0 & 1\end{array} \right)$,  $
Z_2(\theta)= \left( \begin{array}{cccc}
1 & 0 & 0\\
0 & 1 & 0\\
0 & 0 & e^{i\theta}\end{array} \right)$,

\noindent and the qutrit Hadamard gate $H_3=\frac{1}{\sqrt{3}} \left( \begin{array}{cccc}
1 & 1 & 1\\
1 & e^{i\frac{2\pi}{3}} & e^{-i\frac{2\pi}{3}}\\
1 & e^{-i\frac{2\pi}{3}} & e^{i\frac{2\pi}{3}}\end{array} \right)$.
\vspace{0.5em}

Using the above gate definitions, the NAE, NPE and NCE encoding schemes for qutrits (Eq. \ref{eq_NAE}-\ref{eq_NCE} with $d=3$) can be written and implemented as follows:
\begin{flalign}
    \ket{\psi_\mathrm{NAE}} & = \cos{x_0}\ket{0}+\sin{x_0}\cos{x_1}\ket{1}+\sin{x_0}\sin{x_1}\ket{2}, && 
    \label{eq_d3NAE} \\
\nonumber & \Qcircuit @C=1.2em @R=1em {
\ket{0} & & \gate{R_{y01}(2x_0)} & \gate{R_{y12}(2x_1)} & \qw 
}
\\
   \ket{\psi_\mathrm{NPE}} &= \frac{1}{\sqrt{3}}\left(\ket{0}+e^{ix_0}\ket{1}+e^{ix_1}\ket{2}\right), && 
   \label{eq_d3NPE} \\
\nonumber & \Qcircuit @C=1.2em @R=1em {
\ket{0} & & \gate{H_3} & \gate{Z_1(x_0)} & \gate{Z_2(x_1)} & \qw 
}
\\
    \ket{\psi_\mathrm{NCE}}&= \cos{x_0}\ket{0}+e^{ix_2}\sin{x_0}\cos{x_1}\ket{1}+e^{ix_3}\sin{x_0}\sin{x_1}\ket{2}. &&
    \label{eq_d3NCE} \\
\nonumber & \Qcircuit @C=1.2em @R=1em {
\ket{0} & & \gate{R_{y01}(2x_0)} & \gate{R_{y12}(2x_1)} & \gate{Z_1(x_2)} & \gate{Z_2(x_3)} & \qw 
}
\end{flalign}

\noindent For the two-qutrit gate, we use a parameter-free SUM gate, which is a straightforward generalization of the CNOT gate:
\begin{flalign*}
\text{SUM} = \left( \begin{array}{cccccccccc}
1 & 0 & 0 & 0 & 0 & 0 & 0 & 0 & 0\\
0 & 1 & 0 & 0 & 0 & 0 & 0 & 0 & 0\\
0 & 0 & 1 & 0 & 0 & 0 & 0 & 0 & 0\\
0 & 0 & 0 & 0 & 0 & 1 & 0 & 0 & 0\\
0 & 0 & 0 & 1 & 0 & 0 & 0 & 0 & 0\\
0 & 0 & 0 & 0 & 1 & 0 & 0 & 0 & 0\\
0 & 0 & 0 & 0 & 0 & 0 & 0 & 1 & 0\\
0 & 0 & 0 & 0 & 0 & 0 & 0 & 0 & 1\\
0 & 0 & 0 & 0 & 0 & 0 & 1 & 0 & 0 \end{array} \right). &&
\end{flalign*}

While the above definitions of single-qutrit rotations are straightforward, when considering hardware implementations, one may choose to use different definitions to suit the native gates that can be implemented by the hardware easily. In our implementation of the one-qutrit classifier on the transmon device, we chose the following definitions of $X'_{01}$ and $X'_{12}$:

\noindent $X'_{01}(\theta) = H_1 Z_1(\theta) H_1 = e^{i\frac{\theta}{2}}\begin{pmatrix}
\cos{\frac{\theta}{2}} & -i\sin{\frac{\theta}{2}} & 0 \\ -i\sin{\frac{\theta}{2}} & \cos{\frac{\theta}{2}} & 0\\
0 & 0 & e^{-i\frac{\theta}{2}}
\end{pmatrix}$, $X'_{12}(\theta) = H_2 Z_2(\theta) H_2 = e^{i\frac{\theta}{2}}\begin{pmatrix}
e^{-i\frac{\theta}{2}} & 0 & 0 \\
0 &\cos{\frac{\theta}{2}} & -i\sin{\frac{\theta}{2}} \\ 
0 &-i\sin{\frac{\theta}{2}} & \cos{\frac{\theta}{2}}
\end{pmatrix}$, 

\vspace{0.5em}
\noindent where $H_1= Z_1(\pi)R_{y01}(-\pi/2)=\begin{pmatrix}
\frac{1}{\sqrt{2}} & \frac{1}{\sqrt{2}} & 0 \\
\frac{1}{\sqrt{2}} & -\frac{1}{\sqrt{2}} & 0 \\ 
0 & 0 & 1
\end{pmatrix}$, $H_2= Z_2(\pi)R_{y12}(-\pi/2)=\begin{pmatrix}
1 & 0 & 0 \\
0 & \frac{1}{\sqrt{2}} & \frac{1}{\sqrt{2}}\\ 
0 & \frac{1}{\sqrt{2}} & -\frac{1}{\sqrt{2}}
\end{pmatrix}$.

\vspace{0.5em}
\noindent The resulting $X'_{01}(\theta)$ and $X'_{12}(\theta)$ differ from the $R_{x01}$ and $R_{x12}$ derived from Gell-Mann matrices by local phases. We choose these gates because they can be implemented on our transmon device with common microwave pulses, and have two eigenvalues instead of three, which enables us to apply a rotosolve-like optimization algorithm ~\cite{Ostaszewski2021structure,Nakanishi2020Sequential} to optimize for the gate parameters. 

\subsection{Quantum circuits for general qudit encodings}
\label{sesc:SMdataencodings}
\begin{figure*}[htbp]
\centering
\includegraphics[width=0.98\textwidth]{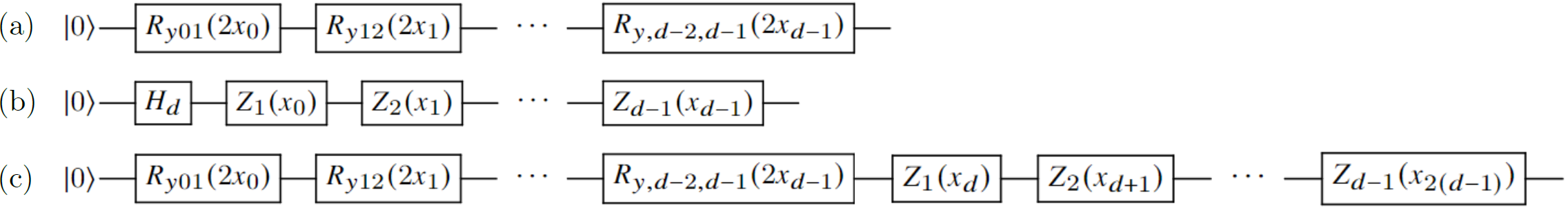}
\caption{Qudit circuits to implement the (a) NAE, (b) NPE and (c) NCE encoding schemes.}
\label{f_qudit_encoding}
\end{figure*}

The quantum circuit implementations of the qudit NAE, NPE and NCE encoding schemes (Eqs. \ref{eq_NAE}-\ref{eq_NCE}) are shown in Fig. \ref{f_qudit_encoding} (a-c), where $R_{yuv}(\theta)$ is the Y-rotation of angle $\theta$ in the subspace $\{\ket{u}, \ket{v}\}$, $H_d$ is a $d$-dimensional Hadamard gate, and $Z_{u}(\theta)$ is a phase of $e^{i\theta}$ applied to the $\ket{u}$ state.

\subsection{Decomposition of an arbitrary SU(3) unitary into the $R_L$ operation}
\label{Appx_decomposition}
We consider an arbitrary SU(3) unitary $U=\begin{pmatrix}
a_{11} & a_{12} & a_{13} \\
a_{21} & a_{22} & a_{23} \\
a_{31} & a_{32} & a_{33} 
\end{pmatrix}$, where $UU^\dagger=I$. We define the following transition unitaries $T_{12}, T_{23}$, and by left and right multiplications onto $U$ we aim to transform $U$ into a diagonal matrix $D$ \cite{Clements2016}:
\begin{flalign}
\nonumber &T_{12}(\beta,\phi)=\begin{pmatrix}
e^{i\phi}\cos{\beta} & -\sin{\beta} & 0\\
e^{i\phi}\sin{\beta} & \cos{\beta} & 0\\
0 & 0 & 1
\end{pmatrix}, \quad
T_{23}(\beta,\phi)=\begin{pmatrix}
1 & 0 & 0\\
0 & e^{i\phi}\cos{\beta} & -\sin{\beta}\\
0 & e^{i\phi}\sin{\beta} & \cos{\beta}
\end{pmatrix}, &&\\
&D = T_{23}(\beta_3, \phi_3) \cdot T_{12}(\beta_2, \phi_2) \cdot U \cdot T_{12}^{-1}(\beta_1, \phi_1). &&
\end{flalign}

\noindent By solving for the requirement that $D$ is diagonal, one set of real solutions is:
\begin{flalign*}
&\beta_3 = \text{arccot}\left|\frac{(a_{21}a_{12}-a_{11}a_{22})|a_{31}a_{12}-a_{11}a_{32}|}{(-a_{31}a_{12}+a_{11}a_{32})\sqrt{|a_{11}a_{32}-a_{31}a_{12}|^2+|a_{31}a_{22}-a_{21}a_{32}|^2}}\right| , &&\\
&\phi_3 = -i \log \frac{(a_{11}a_{32}-a_{31}a_{12})|a_{21}a_{12}-a_{11}a_{22}|}{(a_{21}a_{12}-a_{11}a_{22})|a_{31}a_{12}-a_{11}a_{32}|} , &&\\
&\beta_2 = \text{arccot}\left| \frac{-a_{31}a_{12}+a_{11}a_{32}}{a_{31}a_{22}-a_{21}a_{32}} \right| , &&\\
&\phi_2 = -i \log \frac{(a_{21}a_{32}-a_{31}a_{22})|a_{31}a_{12}-a_{11}a_{32}|}{(a_{31}a_{12}-a_{11}a_{32})|a_{31}a_{22}-a_{21}a_{32}|} , &&\\
&\beta_1 = \text{arctan}\left| \frac{a_{31}}{a_{32}} \right| , &&\\
&\phi_1 = i \log \frac{a_{32}|a_{31}|}{a_{31}|a_{32}|} . &&
\end{flalign*}

Note that the solution is not unique, and here we only give one example. Now we bring $U$ to the left side of the equation, and all other terms to the right side, and obtain
\begin{flalign}
U = T_{12}^{-1}(\beta_2, \phi_2) \cdot T_{23}^{-1}(\beta_3, \phi_3) \cdot D \cdot T_{12}(\beta_1, \phi_1). &&
\label{appx_eq_T12T23_decompose}
\end{flalign}

\noindent Denoting the $j$-th diagonal element of $D$ as $e^{i\gamma_{j}}$, we can further decompose $U$ into the $R_L$ sequence described in the main text using $R_x$ and $R_z$ (Eq. \ref{eq_RL_theoretical}), up to a global phase, as:
\begin{flalign}
U = e^{i\frac{1}{3}(\gamma_1+\gamma_2+\gamma_3+\phi_1-\phi_2-\phi_3)} R_{z01}(\theta_8) \cdot R_{x01}(\theta_7) \cdot R_{z12}(\theta_6) \cdot R_{x12}(\theta_5) \cdot R_{z12}(\theta_4) \cdot R_{z01}(\theta_3) \cdot R_{x01}(\theta_2) \cdot R_{z01}(\theta_1), && 
\end{flalign}
where $\theta_8=\phi_2+\frac{\pi}{2}$, $\theta_7=-2\beta_2$, $\theta_6=\frac{1}{2}\phi_2+\phi_3+\frac{3}{4}\pi$, $\theta_5=-2\beta_3$, $\theta_4=\frac{1}{6}(-4\gamma_1-4\gamma_2+8\gamma_3-4\phi_1+\phi_2-2\phi_3)-\frac{3}{4}\pi$, $\theta_3=\frac{1}{3}(-4\gamma_1+2\gamma_2+2\gamma_3-\phi_1+\phi_2-2\phi_3)$, $\theta_2=2\beta_1$, $\theta_1=-\frac{\pi}{2}-\phi_1$.

\subsection{Equivalence of circuits used in the simulation and in the hardware implementation}
\label{Appx_equivalence}
In our hardware implementation, the encoding circuit using the gate sequence $X'_{01} X'_{12} Z_1 Z_2$ can be proven to be equivalent to the qutrit NCE encoding using the gate sequence $R_{y01} R_{y12} Z_1 Z_2$ (Eq. \ref{eq_d3NCE}), given that the encoding is trained according to the optimized encoding method described in Sec. \ref{method_encoding_training} in the main text:
\begin{flalign}
\nonumber
\ket{\psi}&=Z_2(x_3)\cdot Z_1(x_2) \cdot X'_{12}(x_1) \cdot X'_{01}(x_0) \cdot \ket{0} && \\
&=  e^{i\frac{x_0}{2}} \left( \cos{\frac{x_0}{2}}\ket{0} + e^{i (\frac{x_1}{2}+x_2+\frac{3\pi}{2})} \sin{\frac{x_0}{2}} \cos{\frac{x_1}{2}}\ket{1} + e^{\frac{x_1}{2}+x_3+\pi} \sin{\frac{x_0}{2}}\sin{\frac{x_1}{2}}\ket{2} \right), &&
\label{appx_eq_hardware_encoding}
\end{flalign}
which is equivalent to Eq. \ref{eq_d3NCE} up to a global phase and linear combinations of parameters $x_{0\cdots3}$. When we train the encoding to have $\boldsymbol{\phi}=\mathbf{Wx}+\mathbf{b}$ replacing $\mathbf{x}$ to be fed into the encoding circuit, the changes in parameters are absorbed into the training of $\mathbf{W}$ and $\mathbf{b}$.

For the implementation of the classifier circuit, we show that we can decompose the arbitrary unitary $U$ in Eq. \ref{appx_eq_T12T23_decompose} into the a sequence of $X'$ and $Z$ up to a global phase as well:
\begin{flalign}
U = e^{i(\gamma_1-\beta_1+\beta_2+\phi_1-\phi_2)}Z_{1}(\theta_8) \cdot X'_{01}(\theta_7) \cdot Z_{2}(\theta_6) \cdot X'_{12}(\theta_5) \cdot Z_{2}(\theta_4) \cdot Z_{1}(\theta_3) \cdot X'_{01}(\theta_2) \cdot Z_{1}(\theta_1), &&
\end{flalign}
where $\theta_8=\phi_2+\frac{\pi}{2}$, $\theta_7=-2\beta_2$, $\theta_6=-\beta_2+\phi_2+\phi_3+\pi$, $\theta_5=-2\beta_3$, $\theta_4=-\gamma_1+\gamma_3+\beta_1+\beta_3-\phi_1-\phi_3-\pi$, $\theta_3=-\gamma_1+\gamma_2+\beta_3-\phi_3$, $\theta_2=2\beta_1$, $\theta_1=-\phi_1-\frac{\pi}{2}$.

Note that in the experiment we immediately measure the qutrit in the Z-basis, so the leftmost $Z_1$ has no effect to the measurement and can be omitted. Therefore, we add a redundant $Z_1$ in between the leftmost $X'_{01}$ and $Z_2$ to compensate for the reduction in the number of gate parameters. This results in the implemented $R'_L$ sequence is as presented in the main text in Eq. \ref{eq_RL_hardware}.

\section{Optimization details of simulations}
\label{AppxOptimization}
We start the simulation by choosing a particular setup of task, including the choices of datasets, number of qudits and encoding, as well as choosing a permutation and assignment of data features into the encoding gates. In the case of using the optimized encoding method, the permutation does not apply.

For a setup we run the following process 50 times. We first set a random number generating seed. Then we randomly split the dataset into training and testing sets with a ratio of 2:1. For each data point $\mathbf{x}$ in the training set and its associated class label $y \in \{0, 1, 2\}$, we encode the data into a state vector according to Eqs. \ref{eq_NAE}-\ref{eq_NCE} with $d$=2 or $d$=3. If using the optimized encoding method, we train for a $\mathbf{W}$ and $\mathbf{b}$ and use them to transform the data according to the method described in Sec. \ref{method_encoding_training}, otherwise we apply the chosen data permutation to obtain an initial state vector. After encoding, we randomly initialize the classifier circuit gate angle parameters. We run the simulation of the circuit on the encoded state vector, and obtain a final state vector. A loss value $\mathcal{L}$ for this data point is then calculated with $\mathcal{L}=(1-P(\ket{y}))^2$, where $P(\ket{y})$ is the probability of the $\ket{y}$ state upon measurement. For one qutrit, $\ket{y}$ corresponds to $\ket{0}$, $\ket{1}$ or $\ket{2}$; for two qutrits, $\ket{y}$ corresponds to $\ket{0}$, $\ket{1}$ or $\ket{2}$ of the second qutrit, and the first qutrit's state is ignored; for two qubits, $\ket{y}$ corresponds to $\ket{00}$, $\ket{01}$ or $\ket{10}$.
Then these individual loss values of all data points in the training set are summed to obtain the total loss, as given in Eq. \ref{eq_LSE}.
There is a special case for one qubit, because one qubit only has two states. For this case we define the loss function for a single data point as $\mathcal{L}=\left\{
		\begin{array}{lll}
			P(\ket{0}) & \mbox{if } y=0 \\
			|P(\ket{0}) - 0.5| & \mbox{if } y=1 \\
			1 - P(\ket{0}) & \mbox{if } y=2.
		\end{array}
	\right.$

This total loss value is minimized using the L-BFGS-B optimization algorithm with default hyper-parameters. After the optimization converges, the final parameters are saved and the process is carried out once more on the testing set. For each testing data point, the predicted class label is inferred based on the probabilities of the final state. For one-qubit circuits, the probability of the $\ket{0}$ being in which of the intervals [0,0.33], [0.33,0.67] or [0.67,1] gives the class label. For two-qubit, one-qutrit and two-qutrit circuits, which state from \{$\ket{00},\ket{01},\ket{02}$\}, \{$\ket{0},\ket{1},\ket{2}$\} or \{$\ket{0},\ket{1},\ket{2}$ for the second qutrit\} has the highest probability gives the class label. Predicted class labels are compared against true class labels, and a testing accuracy value is calculated. If the testing accuracy is below 80\%, we re-run the optimization process again for the same setup and same training dataset but using a different random number seed for optimisation, so we can avoid the occasional situation where the optimizer is trapped in a local minimum. We keep re-running the optimization process until either the testing accuracy is above 80\%, or we have re-run for 10 times, at which point we conclude the result is not from a local minimum and accept the result. 

For the 50 random initializations, we obtain 50 testing accuracy values. Thereby, a mean testing accuracy value is obtained by averaging over these 50 values. Finally, for all the 24 or 12 permutations associated with a qudit encoding category, the 24 or 12 mean testing accuracy values are collected and drawn as a box plot, and the mean accuracy obtained with the optimized encoding method is drawn as a star, as shown in Fig. \ref{f_main} in the main text.

\end{document}